\documentclass[preprints,review,accept,oneauthor,pdflatex]{mdpi}

\firstpage{1} 
\pubvolume{xx}
\issuenum{1}
\articlenumber{5}
\pubyear{2020}
\copyrightyear{2020}
\history{Received: date; Accepted: date; Published: date}

\Title{Star Formation in the Ultraviolet}

\Author{Jorick S. Vink $^{1}$\orcidA{}}

\AuthorNames{Firstname Lastname, Firstname Lastname and Firstname Lastname}

\address{%
$^{1}$ \quad Armagh Observatory and Planetarium, College Hill, BT61 9DG, Armagh, UK; jorick.vink@armagh.ac.uk\\}

\abstract{
With the launch of JWST and the upcoming installation of extremely large telescopes, the first galaxies in our Universe will finally be revealed. Their light will be dominated by massive stars, which peak in in the ultra-violet (UV) part of the electromagnetic spectrum. Star formation is the key driver of the evolution of our Universe. At young ages, within 10 Million years, both high and low mass stars generate complex UV emission processes which are poorly understood yet are vital for interpreting high red-shift line emission. For these reasons, the Hubble Space Telescope (HST) will devote 1000 orbits to obtaining a UV Legacy Library of Young Stars as Essential Standards (ULLYSES). The purpose of this Overview is to outline the basic physical principles driving UV emission processes from local (within 100 parsecs of) star formation, ranging from huge star-forming complexes containing hundreds of massive and very-massive stars (VMS), such as 30 Doradus (the Tarantula Nebula) in the neighboring Magellanic Clouds (only 50 kpc away), to galaxies near and far, out to the epoch of Cosmic Reionization.}


\begin{document}

\section{Introduction}
\begin{figure}
\centering
\includegraphics[width=13cm]{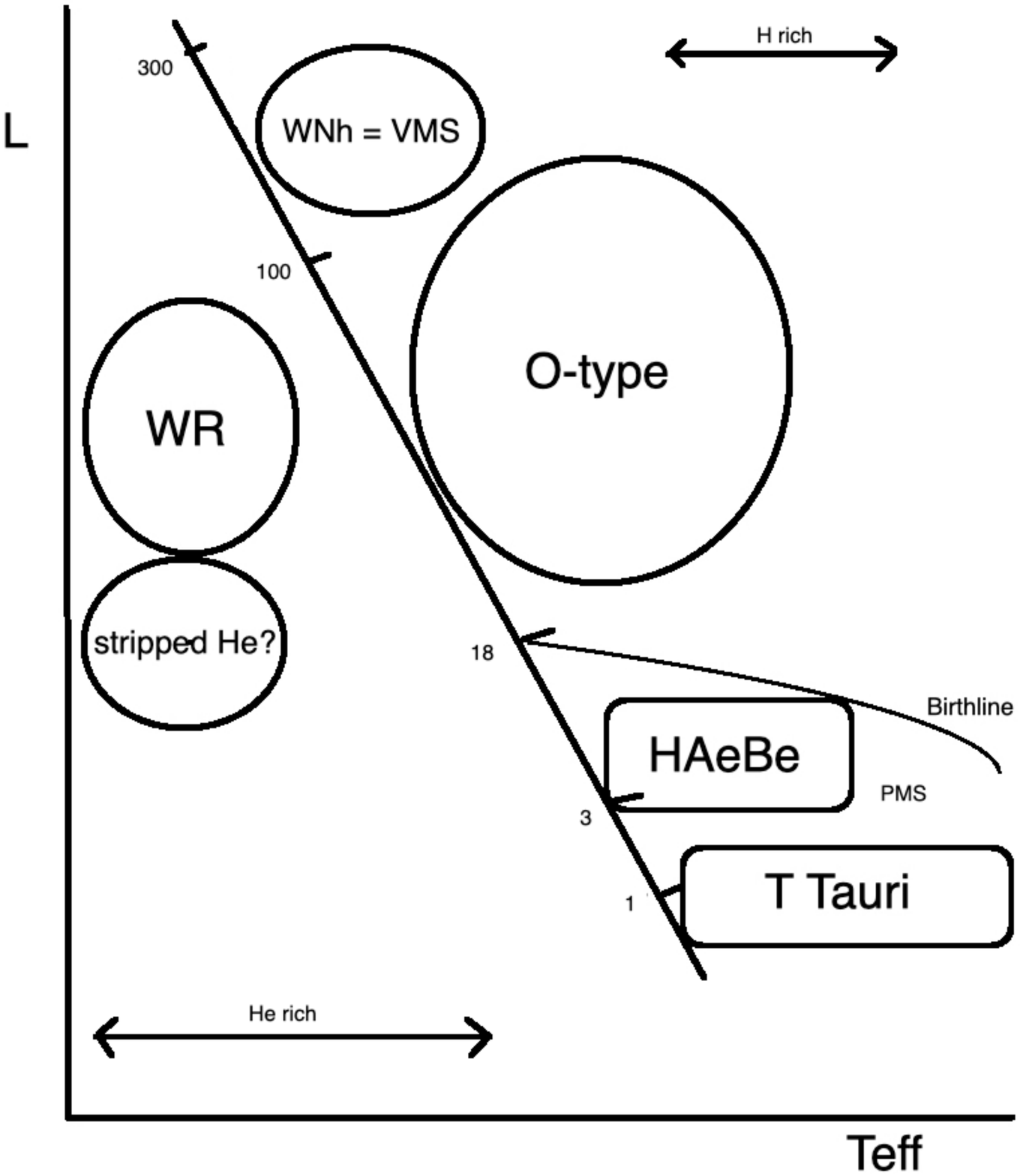}
\caption{Schematic HRD including groups of objects discussed in this overview. Indicative masses are shown to the left of the main sequence. The groups of PMS that we discuss are T Tauri and Herbig Ae/Be stars. The He-rich stars are the classical WR stars and the potential stripped He stars. Massive stars do not have a visible PMS phase, i.e. they are above the birthline. The boundary between canonical massive O stars and VMS is around 100\,M$_{\odot}$ }
\label{hrd}
\end{figure}

Star formation in the Universe has taken place over the vast majority of its existence. The First Stars, some couple of hundred million years after the Big Bang are thought to be massive due to their pristine chemistry, implying less efficient cooling during their formation (e.g. Abel et al. 2000, Bromm et al. 1999). 
Stars in today's Universe mostly have masses like our Sun, which are thought to go through a so-called T Tauri pre-main sequence (PMS) phase (see Fig.\,\ref{hrd}).

In this Editorial overview, and the Special Volume in front of you, we show how the ultraviolet (UV) part of the electromagnetic spectrum is arguably the most important wavelength range to obtain key physical observables for both current PMSs, as well as earlier generations of more massive stars at high redshift.
In terms of Cosmic chronology it would make sense to start our discussion with the First Stars at zero metallicity, before discussing current-day T Tauri and Herbig Ae/Be stars at solar metallicities. However, from an observational perspective it is more pragmatic to start the discussion with PMSs in our local neighborhood. We then gradually increase distances to parts of our Milky Way that also contain intermediate mass Herbig Ae/Be stars (within a few hundred parsecs), before we ultimately discuss the population of the most massive stars in the Local Universe at kilo-parsec scales.

Based on these considerations, this editorial overview has the following structure:

\begin{enumerate}[leftmargin=*,labelsep=4.9mm]
\item	Overview: Star Formation in the Ultraviolet -- Jorick S. Vink 
\item	The UV Perspective of Low-Mass Star Formation -- Schneider, G\"unther \& France
\item	On the mass accretion rates of Herbig Ae/Be stars -- Ignacio Mendigutia
\item UV Spectroscopy of massive stars -- John Hillier
\item Massive Star Formation in the Ultraviolet observed with HST -- Claus Leitherer
\item Applications of Stellar Population Synthesis in the Distant Universe -- Elizabeth Stanway
\end{enumerate}

In chapter 2, Schneider et al.\,discuss the formation of solar-mass stars. Although the mass-accretion rates of T Tauri stars are oftentimes determined from optical line emission, such as H$\alpha$, all these diagnostics ultimately find their roots in the accretion shocks revealed by the UV part of the spectrum. 
In Chapter 3, Ignacio Mendigutia discusses a possible extension of the successful
magneto-spheric accretion model applied to T Tauri stars to the higher mass regime
of the 2-18\,$M_{\odot }$ Herbig Ae/Be stars. His conclusion is that the model may indeed be applicable to Herbig Ae stars up to a few solar masses, but that alternatives like the boundary-layer (BL) model may need to be considered
for the more massive Herbig Be stars.

Yet more massive stars remain above the birthline (see Fig.\,\ref{hrd}) during their PMS phase, and the most massive stars over 18 $M_{\odot}$ only become optically visible after hydrogen (H) burning has started. The objects eventually appear either as either O-type stars (above 18\,M$_{\odot}$), or very massive,  
H-rich Wolf-Rayet (WR) stars (over approx. 100\,$M_{\odot}$). 
Optically visible O and WR stars however have their key diagnostics at UV wavelengths, and their spectra are extensively discussed in Chapter 4 by John Hillier.

The further we look into the Universe, the more we look back in
time. Therefore, the metallicity ($Z$) – on average – is dropping. This
has major consequences, not only on the driving of the winds of O and WR stars, but potentially also
on their evolution and diagnostics.
Moreover, it has become clear that roughly half the massive stars are part of a close binary system (e.g. Sana et al. 2013), further complicating the interpretation of integrated light in clusters and galaxies (discussed by Claus Leitherer in Chapter 5), and into the more distant Universe (Elisabeth Stanway; Chapter 6).

Reflecting the order of the Review chapters in this Special Volume, this Editorial overview will start with the role of the UV for mass-accretion studies of PMSs in the Local Universe. I  then move on to discussing the physics of O and 
 WR stars in low $Z$ environments, before providing an outlook of how to detect the First Stars in the Universe. 

\section{T Tauri and Herbig Ae/Be stars}

For low-mass T Tauri stars there is a well established paradigm involving magneto-spheric accretion (see Fig.\,2 in Chapter 2 by Schneider et al. and similar versions by e.g. Hartmann et al. 2016). The key physics is that due to the presence of a strong kilo-Gauss (approximate) dipole field the inner accretion disk is disrupted, and gaseous material is funnelled along magnetic loops, shocking onto the stellar surface, producing UV radiation (see e.g. Fig.\,5 of Schneider et al. and also Calvet \& Gullbring 1998). 

Half of the HST orbits of ULLYSES (Ultraviolet Legacy Library of Young Stars as Essential Standards) are dedicated to better understanding the physics of mass-accretion in the UV. Some HST orbits are dedicated to spectroscopic monitoring, as rotational modulation has a key effect on mass-accretion rate diagnostics, such as H$\alpha$. The low-mass component of ULLYSES should thus help to unravel the rotationally modulated complex spectral diagnostics. 

Time variability studies of line emission of PMS Herbig Ae stars have also shown rotational modulation to be a decisive factor. 
In Chapter 3, Ignacio Mendigutia highlights a range of additional evidence in favour of the T Tauri model of magnetospheric accretion to be applicable in the Herbig Ae range. Furthermore, it is argued that the more massive Herbig Be stars are subjected to a different mode of accretion, possibly involving a boundary layer. 

Independent of the specific mode of accretion along the stellar mass sequence, it is clear that the UV will play a vital role in deciding the modes of accretion, and for deriving accurate mass-accretion rates as a function of stellar mass. 
A relatively unexplored physical dimension in the PMS range is that of $Z$. This is mostly due to the fact that PMS stars are faint and spectroscopy of PMSs in low-$Z$ environments is only slowly starting to take off, sometimes in the form of "extremely low resolution spectroscopic" i.e. {\it narrow-band photometric} methods. At the moment it is still under debate whether the mass-accretion rates of PMS stars depend on $Z$, or not (de Marchi et al. 2011 vs. Kalari \& Vink 2015).

\section{Massive and very massive stars}

Massive stars dominate the light of star-forming galaxies and are also thought to dominate the light from the First Galaxies. In certain circumstances it might even be the case that most of the stellar mass is locked up in massive stars, leading to a top-heavy initial mass function (IMF; Schneider et al. 2018). whether the IMF of the First Stars
was top-heavy is still under debate, but it is considered likely on theoretical grounds (e.g. Abel et al. 2000; Bromm et al. 1999; Haemmerl\'e
et al. 2020).

\begin{figure}
\centering
\includegraphics[width=15 cm]{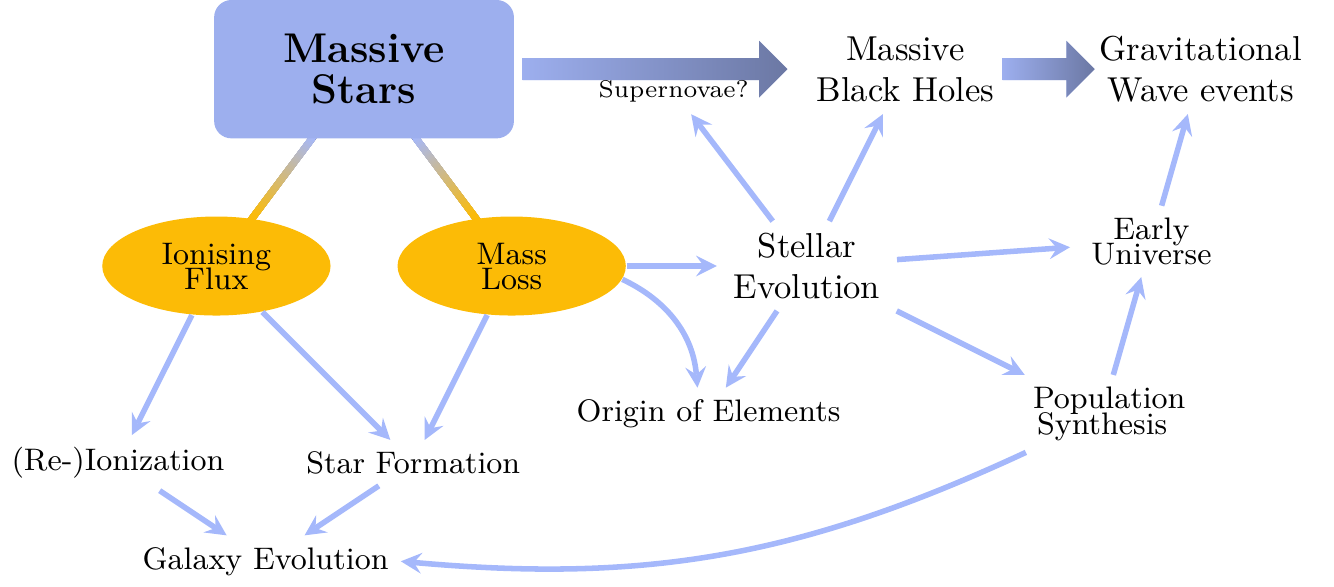}
\caption{Overview of how massive stars, and in particular the mass-loss rates and ionising fluxes, determined in the UV, impact many adjacent areas of Astrophysics. Image credit: Andreas Sander.}
\label{andreas}
\end{figure}

Even in Today's Universe massive stars are key cornerstones to many aspects of Astrophysics (see Fig.\,\ref{andreas} for an overview).
In order to understand gravitational wave (GW) events as a function of cosmic time and $Z$, we need to understand the evolution \& winds of
massive stars at low $Z$. For our understanding of stellar feedback and 
He {\sc ii} emission at high redshift we need to be able to predict the 
ionising and UV radiation of massive stars which is determined by their effective temperatures and mass-loss rates. In order to properly predict the spectrum of a massive star it is pivotal to apply non-LTE (local thermodynamic equilibrium) radiative transfer models of spherical expanding atmospheres, as described in Chapter 4 by John Hillier, and compare these models to large samples of spectra at UV (ULLYSES contains 500 HST orbits on massive OB and WR stars) and optical wavelength ranges.
This is needed given the 
extreme observational uncertainties associated with mass-loss rates of massive stars below LMC metallicity (e.g. Mokiem et al. 2007; Bouret et al. 2015; Tramper et al. 2015; Evans et al. 2019; Garcia et al. 2019; Ramachandran et al. 2019), and the discovery-space available to diagnose the wind -- and in the case of O-stars also the photospheric -- conditions of low-$Z$ stars. 

The expanding non-LTE wind models are generally able to successfully predict the observed UV and optical spectra of the Galactic massive O and WR stars. However, radiation is not only a probe for the conditions in the atmospheres, but also a key constituent itself! It is the radiation pressure gradient that provides an outward acceleration kick-starting an intense radiation-driven wind.
Such winds leave very specific line diagnostics in the electromagnetic (EM)
spectrum, including P Cygni scattering lines in the UV as well as
(usually broad, but see Sect.\,5) recombination emission lines in the UV
(e.g. He {\sc ii} 1640\AA), optical and infrared regime.

\section{WR stars}

 \begin{figure*}
\begin{center}
 \includegraphics[bb=15 70 485 775,height=1.0\columnwidth,angle=-90]{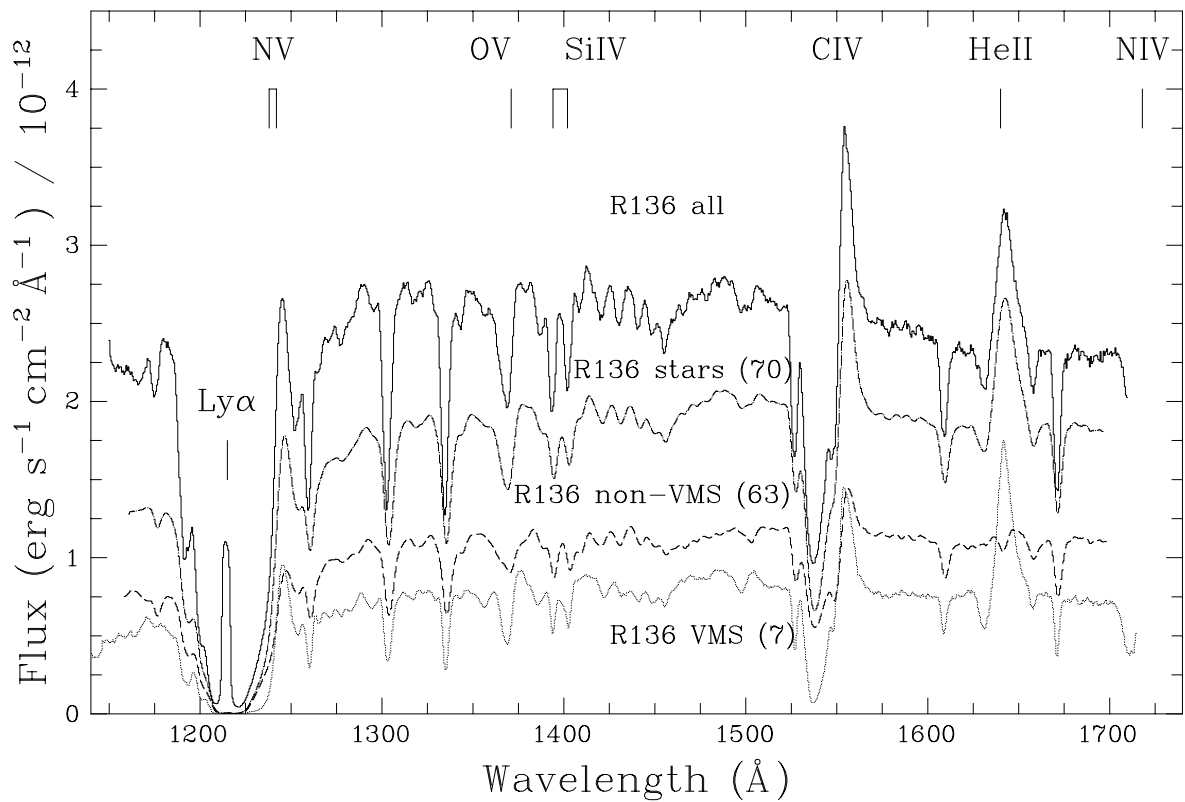} 
 \caption{Integrated HST spectrum of all sources within 0.5 parsec of R136a1 (solid, all stars), the
composite spectrum of all 70 bright stars with    $F_{\rm 1500} \geq 5 \times 10^{-15}$ erg\,s$^{-1}$\,cm$^{-2}$\,\AA$^{-1}$ (dot-dash), comprising 
7 very massive stars (VMS, dotted),
   and the remaining 63 far-UV bright stars (dashed). He\,{\sc ii} 
$\lambda$1640 emission in R136 is totally dominated by VMS. The difference between the `stars' and `all' arises
from the contribution from UV-faint, late-type O
stars and unresolved early B stars. From Crowther et al. (2016).}
\label{r136}
\end{center}
\end{figure*}

In the range of massive stars there are basically 2 types of WR stars: (i)
H-rich VMS, which are VMS that still burn H in their cores, and (ii) classical WR stars (cWRs) that are evolved helium (He) burning stars (see Fig.\,\ref{hrd}). The latter group normally dominates the population, but note that 
the WR phenomenon is a {\it spectroscopic} classification and in principle independent of evolutionary phase (cf. Beals 1940, but see e.g.
Shenar et al. 2020 for a more recent discussion). The reason for the occurrence of emission lines is that the winds have become optically thick ($\tau > 1$) and this means also means that multiple scattering (wind efficiently $\eta > 1$) is a dominant physical process in their winds (Vink \& Gr\"afener 2012). While, the fundamental difference between optically thin O star and
optically thick cWR star winds has been recognized more than two decades
ago, the detailed understanding of cWR winds is still in its early
stages with the differences between the cWR and O star regimes just
starting to become more clear.
 In particular the traditional Castor et al. (1975; CAK) radiation-driven wind parametrization  in terms of force multipliers completely breaks-down in the regime of the cWRs (Sander et al. 2020). 

For H-rich VMS, the transition between normal O stars and H-rich VMS (WNh stars) has already been better mapped in the last decade, both theoretically (Vink et al. 2011) and empirically (Bestenlehner et al. 2014). Due to the proximity to the Eddington limit, these VMS have wind mass-loss rates that are significantly enhanced in comparison
to normal O-star winds, an effect typically not yet accounted for in stellar evolution \& population synthesis models. 
Yet, we know that it is these VMS that are dominant in terms of their 
ionising radiation and kinetic wind input over the canonical O-star population (Crowther et al. 2016; Bestenlehner et al. 2020). Even when the upper-mass limit was still considered to be in the range 120-150 $M_{\odot}$ (Weidner \& Kroupa 2004; Figer 2005; Oey \& Clarke 2005) the most massive stars dominated these quantities (Voss et al. 2009).
With an increase in the upper-mass limit to 200-300 $M_{\odot}$ (Crowther et al. 2010; Martins 2015; Vink 2018; Bestenlehner et al. 2020) this dominance is only expected to grow. 

For the ionising radiation budget and in particular also the expected total He {\sc ii} emission it is relevant 
to realize that locally, such as in the Tarantula Nebula, it is the very massive stars that are the dominant contributors (see Fig.\,\ref{r136} and Doran et al. 2013; Bestenlehner et al. 2020). 
This therefore suggests that VMS also need to be appropriately accounted for in the predictions of the ionizing radiation and additional physical properties of the stellar populations of the Early Universe. 

\section{He II emission at high redshift}
\label{heii-vms}

A fundamental question concerns the origin of the 
sources of the First Light ending the Cosmic Dark Ages and beginning the process of re-ionization. In particular the James Webb Space Telescope (JWST)
and ground-based extremely large telescopes (ELTs) are expected to provide direct access to this critical period via observations of the
first star-forming galaxies at high redshifts ($z \geq 10$).  

\begin{figure}
  \parbox{0.95\textwidth}{\center{\includegraphics[scale=0.75]{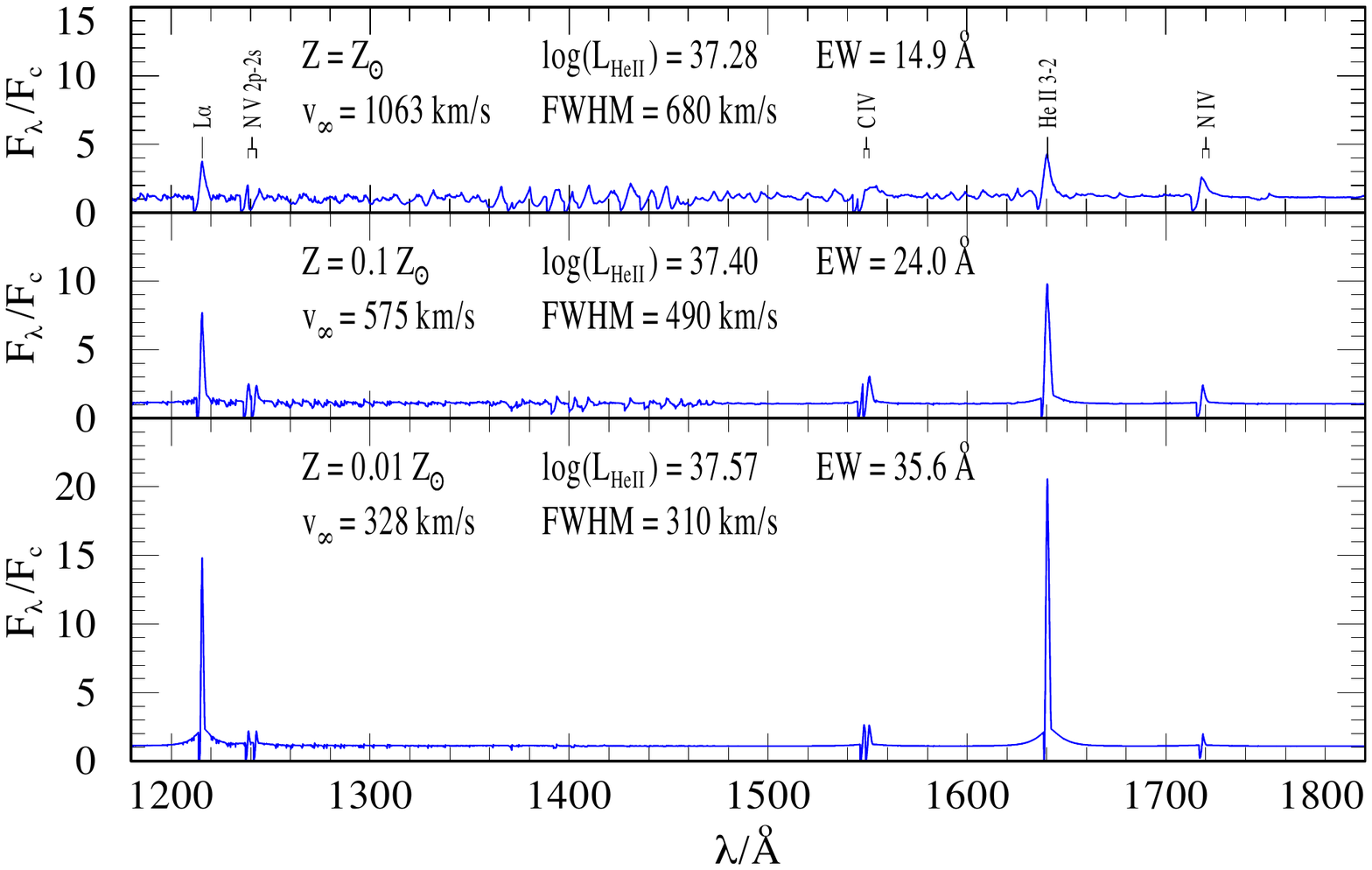}}}
  \caption{Synthetic UV spectra for different metallicities $Z$ from
    wind models for VMS from 
    Gr\"afener \& Vink (2015). The presented
    models are computed for luminosities of $\log(L/L_\odot)=6.3$, a
    stellar temperature $T_\star=45$\,kK, and have very similar
    mass-loss rates ($\sim 1.8 \times 10^{-5}\,M_{\odot}$/yr).}
  \label{fig:graf15}
\end{figure}

Ly$\alpha$ $\lambda$1216 and He {\sc ii} $\lambda$1640 emission are seen as the main indications for stars formed out of pristine gas (the so-called ``Pop III stars''). 
The reason is
that only at extremely low $Z$, massive stars are believed to be hot enough to excite He {\sc ii} in their
surrounding H {\sc ii} regions (Schaerer 2003). The
investigation of star-forming galaxies with redshifted He {\sc ii} emission at
moderate \& high redshifts, which have become increasingly accessible with current ground-based
instrumentation (e.g. Shapley et al. 2003; Eldridge \& Stanway 2012; Steidel et al. 2016), is thus an important preparation for future studies
of the first star-forming galaxies.

In relatively recent studies of moderate redshift ($z$ = 2 - 5) star-forming galaxies 
He {\sc ii}\,$\lambda$1640 emission was found to occur in two modes distinguished by the
width of their emission lines (e.g. Cassata et al. 2013).  Broad emission has
  been attributed to stellar emission from cWR stars, but the origin of narrow He {\sc ii} emission is less obvious. In extra-galactic studies it has generally been attributed to nebular emission excited by a
  population of hot Pop\,III stars formed in pockets of pristine
  gas at moderate redshifts (see also Sobral et al. 2015 for even higher redshift data). 
  
  There are however plausible alternatives to these Pop III postulations. One of them involves the suggestion of {\it stellar}
  emission from VMS at low $Z$ due to a strong but {\it slow} wind.
    Gr\"afener \& Vink (2015) estimated the expected He {\sc ii} line flux
  and equivalent widths (EWs) based on their VMS wind models 
  and Starburst99 (Leitherer et al. 1999) population synthesis models, and compared
  their results with observed star-forming galaxy spectra, finding that the 
  measured He {\sc ii} line strengths and EWs are in line with what is expected for a VMS
  population in one or more young clusters located in
  these galaxies.

    Future high spectral resolution studies could help distinguish between nebular and stellar emission from VMS, taking into account that slower VMS winds yield narrower lines, possibly even below the slow-wind predictions from Gr\"afener \& Vink (2015). Moreover, it is pertinent that new population models are tested on local He {\sc ii} emitting galaxies as well. For instance, for the local very low metallicity analog IZw 18 (with $Z$ below 1/20 $Z_{\odot}$) Kehrig et al (2015) found the spatial extent of the He {\sc ii} emission to be at odds from the location of the massive stellar population, challenging the contribution from stellar He {\sc ii} emission from VMS. It seems clear that in order to explain the full observed range of complex He {\sc ii} line morphologies, both in local He {\sc ii} emitting galaxies (Senchyna et al. 2017; Jaskot et al. 2017; Berg et al. 2018; Erb et al. 2019) and further afield (e.g. Steidel et al. 2016) will require a combined model of  stellar and nebular contributions from a range of sources, involving stellar population and photoionisation modelling.
  
 \section{Stripped stars due to potential binary interaction}
  
  Another potential source of ionising radiation at high redshift, and possibly contributing to cosmic re-ionisation, has recently been put forward: binary-stripped helium stars (G\"otberg et al. 2020; Stanway et al. 2016), which should be located at luminosities just below the classical WR stars (see Fig.\,\ref{hrd}).
  Sander et al. (2020) demonstrated that He star mass-loss rates drop
significantly below a certain luminosity and luminosity-to-mass (L/M) ratio.
This implies that extrapolations from empirical mass-loss recipes applicable to cWR stars (such as Nugis \& Lamers 2000) are inaccurate for stripped He stars, confirming the earlier pilot study results of Vink (2017). 

\begin{figure}[h]
\center 
\includegraphics[width=15cm]{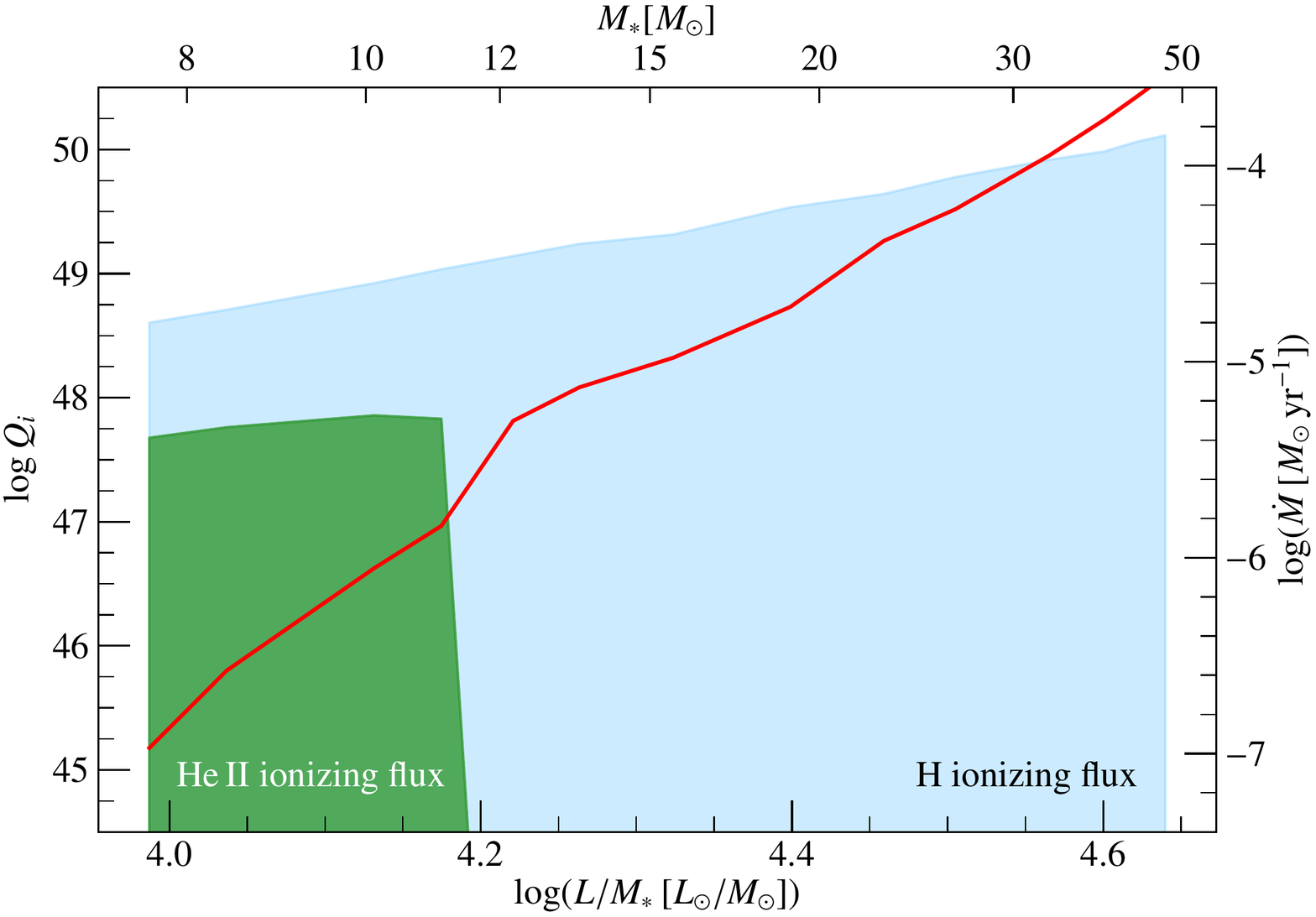}
\caption{Ionising flux and mass-loss predictions from PoWR hydro-dynamical simulations. The red line indicates mass-loss rates (right-hand y-axis) vs. stellar $M$ (top x-axis) and $L/M$ ratio 
(bottom x-axis) for the He main sequence. Ionising fluxes (left-hand y-axis) are indicated with blue/green Boxes. Whilst the H ionising flux (blue box) varies gradually with $M$ and mass-loss rate, the He {\sc ii} ionisation flux (green box) changes abruptly by a factor $>$1000 at a critical $L/M \simeq 4.2$ (from Sander et al. 2020b).}  
\label{sander}
\end{figure}

Figure \ref{sander} showcases state-of-the-art hydro-dynamical PoWR computations -- for L/M ratios from stellar models -- that straddle both the optically thick cWR part as well as the optically thin "stripped star" regime due to Vink (2017).
The figure shows the ionising photon flux for H and He over a wide range of $L/M$ ratios. The most notable aspect is probably that the He {\sc ii} ionisation flux changes abruptly by a factor $>$ 1000 at a critical $L/M$ ratio.

A similar transition between optically thin and thick winds in the H-rich part of the HRD was studied in Vink et al. (2011), but the transition in the He-rich part
of the HRD is only recently being investigated. 
The results from Fig.\,\ref{sander} show that any existing study of stripped stars contributing to the He ionisation of the Universe will necessarily suffer from enormous uncertainties, until we develop a proper understanding of stripped stars via hydro-dynamical stellar atmospheres.

  \section{Summary and Outlook}
  
  In order to predict the feedback from massive stars in star-forming galaxies at low $Z$, we need to better understand the mass-loss rates and associated ionising fluxes from (very) massive stars (VMS), as well as binary-stripped He stars.
 
 In Sect.\,\ref{heii-vms}, the possibility for the existence of VMS with slow winds
at extremely low $Z$ was discussed.  Stellar He {\sc ii} emission from such very early
  VMS generations may become detectable in studies of
  star-forming galaxies at high redshifts with JWST and ELTs.
   The fact that both the {\it stellar} and the {\it nebular} He {\sc ii} emission
  of VMS are still largely neglected in current population synthesis models of massive (single and binary) stars implies that massive progress is urgently required in order to properly interpret the integrated spectra
  of young stellar populations, both nearby \& far-away.
  
  Furthermore, we require an improved understanding of the more canonical massive O and WR stars, especially in low $Z$ galaxies. For these reasons HST has dedicated hundreds of its orbits to build the spectroscopic Legacy survey ULLYSES. This Hubble Atlas will not only provide a fundamental reference data set for UV spectroscopy at low $Z$, but will also be a treasury chest for gaining a greater understanding of the
  winds, evolution, atmospheres, and ionising feedback parameters of massive stars that are urgently required to advance population synthesis modelling in the 21st century.



\acknowledgments{I would like to thank Andreas Sander for constructive comments, and I would like to extend my gratitude to all my other colleagues from the last decade: little would have been achieved without you.}

\reftitle{References}

\end{document}